\RequirePackage{fix-cm}
\documentclass[twocolumn,epjc3]{iopart}  
\RequirePackage{graphicx}
\RequirePackage{xurl}
\RequirePackage[colorlinks,citecolor=blue,urlcolor=blue,linkcolor=blue]{hyperref}
\usepackage{nicefrac}
\RequirePackage{isotope}
\RequirePackage{xcolor}
\RequirePackage{textcomp}
\RequirePackage{xspace}
\usepackage{lineno}
\usepackage{siunitx}
\newcommand{\ts}{$\tau$SPECT}
\newcommand{\tss}{$\tau$SPECT\xspace}
\newcommand{\omegasf}{\omega_{\text{SF}}}
\newcommand{\zdetstorage}{z_{\text{store}}^{\text{det}}}
\newcommand{\zsfstorage}{z_{\text{store}}^{\text{SF}}}
\newcommand{\zsffill}{z_{\text{fill}}^{\text{SF}}}
\newcommand{\zsffillfirst}{z_{\text{SF1}}}
\newcommand{\zsffillsecond}{z_{\text{SF2}}}
\newcommand{\zdetcleaning}{z_\text{clean}^{\text{det}}}
\newcommand{\zdetcounting}{z_\text{count}^{\text{det}}}
\newcommand{\cleaningduration}{t_\text{clean}}
\newcommand{\cleaningstarttime}{t_\text{clean}^{\text{start}}}
\newcommand{\fillingduration}{t_\text{fill}}
\newcommand{\taun}{\ensuremath{\tau_n}\xspace}
\newcommand{\Vud}{\ensuremath{V_\mathrm{ud}}\xspace}
\DeclareSIUnit\bar{bar}

\usepackage{footmisc}

\begin{document}
\title[\ts: A spin-flip loaded magnetic ultracold neutron trap]{\ts: A spin-flip loaded magnetic ultracold neutron trap for a determination of the neutron lifetime}
\author{
J.\,Auler$^1$,
M.\,Engler$^2$,
K.\,Franz$^2$,
J.\,Kahlenberg$^1$,
J.\,Karch$^1$,
N.\,Pfeifer$^1$,
K.\,Roß$^2$,
C.-F.\,Strid$^1$,
N.\,Yazdandoost$^2$,
E.\,Adamek$^1$,
S.\,Kaufmann$^2$,
Ch.\,Schmidt$^1$,
P.\,Bl\"umler$^1$,
M.\,Fertl$^{1}$\footnote{Corresponding author: mfertl@uni-mainz.de},
W.\,Heil$^1$, and
D.\,Ries$^{2,3}$\footnote{Corresponding author: dieter.ries@psi.ch}}
\address{$^1$ Institute of Physics, Johannes Gutenberg University Mainz, 55099 Mainz, Germany}
\address{$^2$
Department of Chemistry - TRIGA site, Johannes Gutenberg University Mainz, 55099 Mainz, Germany} 
\address{$^3$
Paul Scherrer Institut, CH-5232 Villigen PSI, Switzerland}
%
\begin{abstract}
%
The confinement of ultracold neutrons (UCNs) in three-dimensional magnetic field gradient or magneto-gravitational traps allows for a measurement of the free neutron lifetime \taun with superior control over loss channels related to UCNs interacting with material surfaces.
The most precise measurement \taun has been achieved using a magneto-gravitational trap, in which UCN are prevented from escaping at the top of their trap by gravity.
More compact horizontal confinement geometries with variable energy acceptance ranges can be obtained by using steep magnetic field gradients in all spatial directions, generated by combinations of either permanent or (variable) superconducting magnets.
In this paper, we present the first successful implementation of a pulsed spin-flip based loading scheme to fill a three-dimensional magnetic trap with externally produced UCN. 
The measurements with the \tss experiment were performed at the pulsed UCN source of the research reactor TRIGA Mainz. 
The extracted neutron storage time constant of $\tau=\SI{859(16)}{s}$ is compatible with the most precise determinations of \taun.
We report on detailed, but statistically limited,  investigations of major systematic effects influencing the neutron storage time.
The statistical limitations are mitigated by the relocation of the experiment to a stronger UCN source.
\newline
\newline
\noindent{Keywords: Neutron, lifetime, ultracold, UCN, magnetic trap, spin-flip loading}
\end{abstract}
%
\maketitle
\ioptwocol

\section{Introduction}
\label{intro}
The free neutron lifetime is a fundamental parameter for particle and astroparticle physics, as well as for cosmology. 
It is particularly important in the understanding of the production of first light elements during Big Bang Nucleosynthesis and the extraction of the element \Vud of the Cabibbo–Kobayashi–Maskawa matrix without nuclear structure corrections.
To calculate the abundance of first light elements after Big Bang Nucleosynthesis,
the number of relativistic species, baryon-to-photon ratio, neutron-proton mass difference and the neutron lifetime is needed.
This calculated abundance can be compared to measurements of the primordial abundance to search for physics beyond the Standard Model of particle physics \cite{Cyburt2016}. 
Currently, \Vud is determined with the highest precision by superallowed $0^+ \rightarrow 0^+$ decays where nuclear structure corrections are needed \cite{Hardy2020}. 
The neutron as the prototype of a $\beta$-decaying system does not require nuclear structure corrections and therefore offers an independent alternative in the determination of \Vud. 
Here, the only input parameters are the neutron lifetime \taun and the ratio of axial-vector and vector coupling $\lambda = \nicefrac{g_A}{g_V}$ \cite{Maerkisch2019}. 
In order to become competitive to the results obtained from superallowed decays, measurements of \taun with a relative precision of $1\times 10^{-4}$ and of $\lambda$ with a precision of $8 \times 10^{-5}$ or better are required \cite{Hardy2020}.
Two different methods for measuring the free neutron lifetime have been pursued: Beam experiments count protons emerging from a cold neutron beam per unit time and relate this number to the flux of the neutron beam \cite{Yue2013}.
Bottle experiments store ultracold neutrons (UCNs) in traps and count the neutrons which are left in the trap after fixed periods of time \cite{Serebrov2018,Gonzalez2021}. 
The current best measurement for the free neutron lifetime in a beam experiment is $\tau_{\text{beam}}=\SI{887.7\pm2.3}{\second}$ \cite{Yue2013}, and in a magnetic bottle experiment $\tau_{\text{bottle}}=\SI{877.75\pm0.36}{\second}$ \cite{Gonzalez2021}.
These results deviate by $4.3\sigma$. 
In order to understand and resolve this discrepancy, the search for unaccounted systematic effects and physics beyond the Standard Model is ongoing \cite{Byrne2019,Serebrov2021,Dubbers2021}.
Historically, the main source of systematic uncertainties in material bottle type experiments are wall losses that triggered the development of sophisticated mechanisms to characterize them to sufficiently small uncertainty, see e.g. \cite{SEREBROV200572}. 
Over many decades, the often large corrections have in general resulted in neutron lifetime values higher than the currently most precise measurement result \cite{Gonzalez2021}.
State of the art magnetically trapping of UCNs eliminates the largest source of these systematic uncertainties.
Isotopically pure $^{58}\mathrm{Ni}$ has the largest known neutron optical potential (Fermi potential) ($V_F$) among materials of about \SI{335}{\nano\electronvolt}.
Although, other definitions are possible, UCNs are commonly defined to be neutrons with a kinetic energy below this potential.
Neutrons with kinetic energies below \SI{350}{\nano\electronvolt} are called UCNs.
This energy threshold is defined by the largest known neutron optical potential (Fermi potential) ($V_F$) among materials. 
Neutrons with energies equal or below the Fermi potential of a material are reflected under all angles of incidence.
Additionally, neutrons are affected by Earth's gravitational potential ($\Delta E_{\mathrm{grav}} = \SI{102.4}{\nano\electronvolt/\meter}$), on the same order of magnitude as the kinetic energy of UCNs. 
The interaction with magnetic fields depends on the orientation of the neutrons magnetic moment ($\mu = -\SI{60.31}{\nano\electronvolt/\tesla}$)  \cite{Golub1991}.
Neutrons with their magnetic moment oriented antiparallel to a magnetic field are decelerated by an increasing magnetic field. 
Therefore, they are called low field seeking neutrons (LFS). 
Neutrons which are accelerated by an increasing magnetic field are called high field seeking neutrons (HFS).
In a magneto-gravitational trap, the maximum kinetic energy of confinable UCN is set by the achievable vertical height of the experiment.
In a fully magnetic trap, the physical dimensions are determined by the achievable magnetic field gradients.
In particular, predominantly horizontally-oriented trapping volumes with large energy acceptance become feasible.
In-situ production and fully-magnetic storage of LFS UCN in superfluid helium filling a Ioffe-Pritchard trap has been demonstrated earlier \cite{Huffman2000,Bowman2005-rb}.
Here we present the first successful loading a fully magnetic field gradient trap with externally produced UCN using a pulsed spin-flip scheme.
The \tss experiment is located at the pulsed research reactor TRIGA Mainz and uses the ultracold neutron source installed at beam port D \cite{Kahlenberg2017,Bison2017}. 
This paper is first introducing the experimental setup in section~\ref{sec:apparatus}, followed by a discussion of Monte-Carlo simulations of the central trap loading mechanism in section~\ref{sec:filling_sim}. Section~\ref{sec:systematic effects} discusses our current understanding of systematic effects. In section~\ref{sec:procedures} we present the time structure of the experiment employed to record the first UCN storage curves presented in section~\ref{sec:storage_curve}. We summarize our results in section~\ref{sec:summary}.

\section{Experimental setup}
\label{sec:apparatus}
\begin{figure*}[t!]
\centering
 \includegraphics[width=0.9\textwidth]{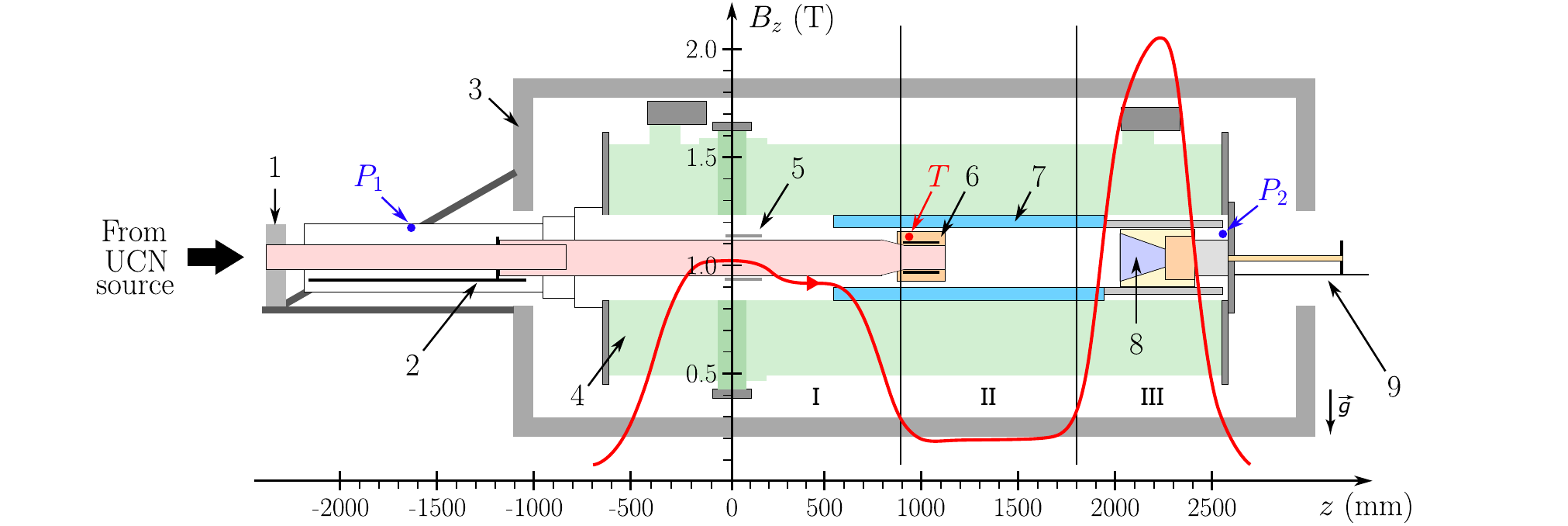}
\caption{Schematic view of the \tss experiment. The cryostat (4) is surrounded by an iron yoke (3) and houses the superconducting coils producing the $B_z$ field (red line), the storage octupole (7) and the neutron detector (8). $B_z$ can be divided into three regions, the field maxima in regions I and III are used for longitudinal confinement of UCNs, the low field in region II maintains the neutron spin polarization. 
UCNs enter the experiment through a shutter (1) and are guided into the magnetic trap by fused silica tubes (shown in pink). The spin flipping unit (SFU, 6) SF2 transforms high-field seeking neutrons into storable low-field seekers. An additional spin flipper SF1 (5) was installed in an upgrade of the experiment. 
Both SFU and detector are movable by translation stages (2) and (9) operated by stepper motors. 
The pressure is monitored at positions $P_1$ and $P_2$, the temperature inside the trap is measured at position \textit{T} within the SFU.
The direction of gravity is given by the vector of gravitational acceleration $\vec{g}$.}
\label{fig:tauSpectDrawing}
\end{figure*}

\tss is the first experiment to store externally produced UCNs in a three-dimensional, magnetic field gradient, and in-vacuum trap. 
A schematic drawing of the experiment is shown in figure\,\ref{fig:tauSpectDrawing}. 
The main components are the confinement magnets, which build up the magnetic trap (Sec.\,\ref{ssec:magnets}), the spin flipping unit (SFU) consisting of an adiabatic fast passage spin flipper (Sec.\,\ref{ssec:afpsf}) surrounded by a Halbach octupole for local field compensation (Sec.\,\ref{ssec:magnets}), the UCN detector (Sec.\,\ref{ssec:detector}) as well as additional components for monitoring purposes (Sec.\,\ref{ssec:monitoring}). 
\subsection{Magnetic trapping fields}
\label{ssec:magnets}
\begin{figure}[t]
\centering
  \includegraphics[width=0.9\columnwidth]{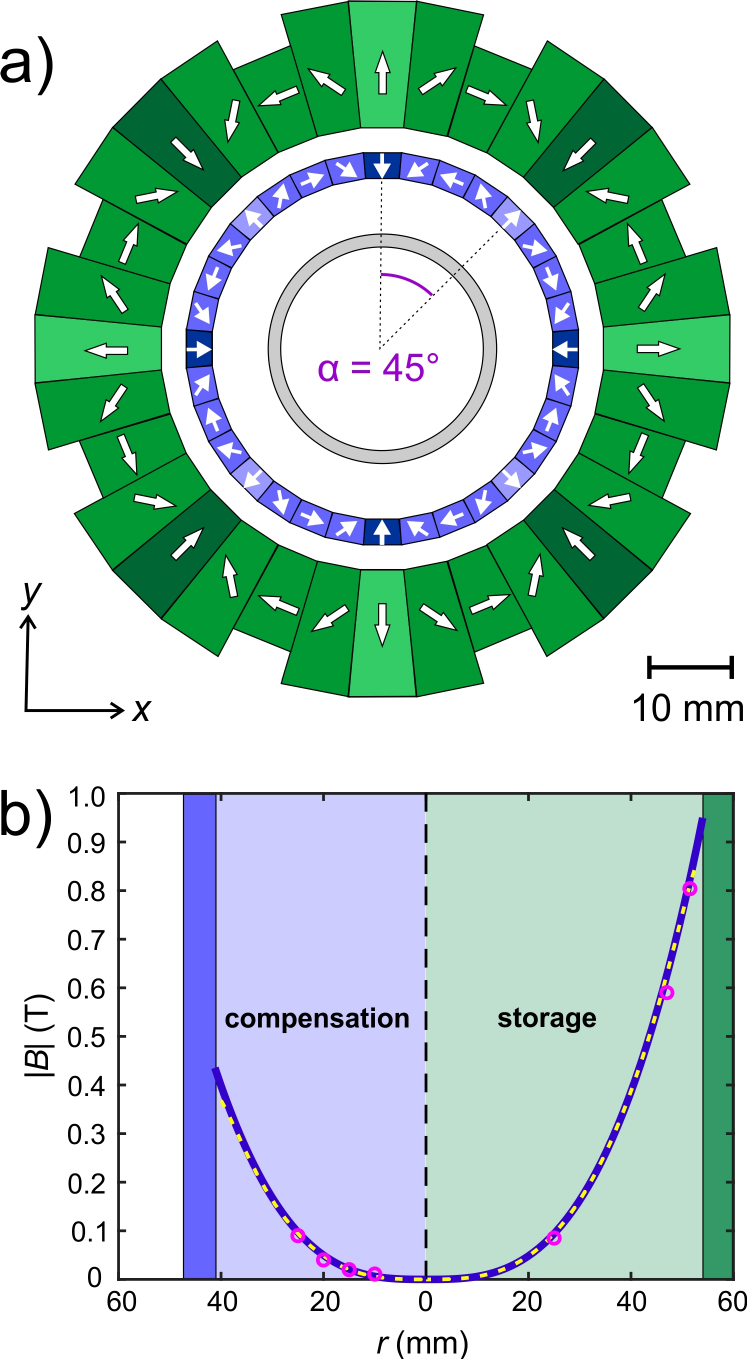}
\caption{
a) Schematic drawing of the two permanent magnet systems: The storage octupole (in green) and the compensation octupole (in blue) are coaxially aligned with the 
neutron guide (gray). The white arrows indicate the magnetization direction of the permanent magnets. The eight poles are accentuated using darker and brighter colors.
b) Comparison of magnitudes of the radial magnetic fields in the center of the storage (right) and compensation (left) octupole magnets. 
The dark blue line shows the result of \ref{eq:HB_length}.
For the storage octupole the arithmetic average ($R_o=\SI{81.75}{\milli\meter}$) of the 24 longer wedges with $R_o=\SI{84.1}{\milli\meter}$ and the eight shorter ones with $R_o=\SI{74.7}{\milli\meter}$ was used. 
The dashed yellow line shows the results from the FEM-simulations, while the magenta circles indicate measurements inside the constructed devices. The solid rectangles mark the position of the magnets. A simulation of the resulting compensated field can be found in figure~\ref{fig:sim_fields}.}
\label{fig:HB_fig}
\end{figure}

\begin{figure}[t]
\centering
  \includegraphics[width=0.99\columnwidth]{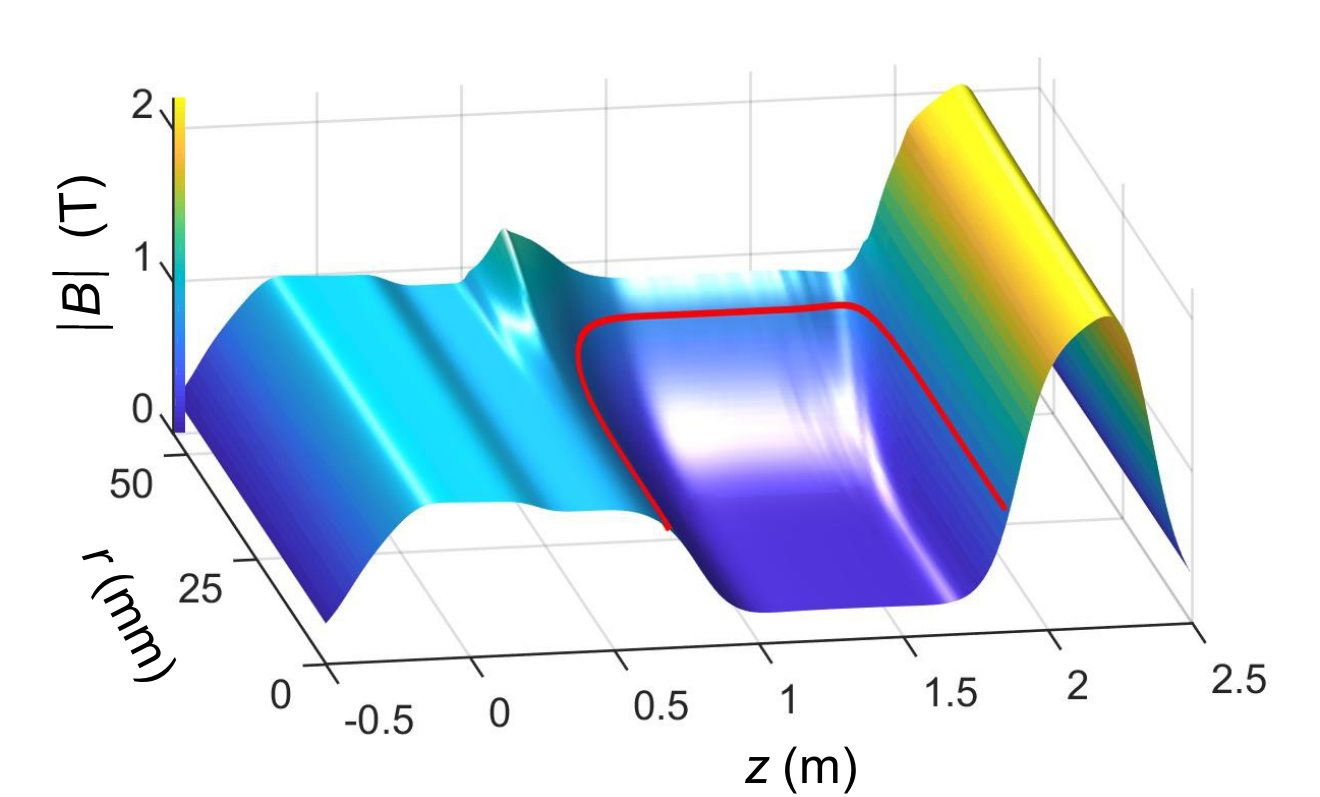}
\caption{Magnetic field distribution forming the three-dimensional magnetic trap. 
UCNs in the low-field seeking spin state are confined within the local minimum of magnetic flux density created by the field superposition of the superconducting coils and the permanent magnet octupole array. The contour line (red) corresponds to a magnetic field of $B=\SI{0.78}{\tesla}$ which translates to a UCN magnetic potential energy of $V_{\text{magn}} = \SI{47.0}{\nano\electronvolt}$.}
\label{fig:3dFields}
\end{figure}

Longitudinal confinement of UCNs in the magnetic trap is achieved by the main magnetic field $B_z$ along the horizontal $z$ axis. Its magnitude along the z-axis is shown by the red line in figure~\ref{fig:tauSpectDrawing}. 
This field is produced by a superconducting multi coil system, which along with its cryostat has been repurposed from the $a$SPECT experiment \cite{Baessler2008,Beck2020}. 
The specific shape of the magnetic field arises from requirements as part of a MAC-E filter\footnote{MAC-E filter stands for '\textbf{M}agnetic \textbf{A}diabatic \textbf{C}ollimation in combination with an \textbf{E}lectrostatic Filter'}\cite{Glueck2005}.
The $B_z$ field can be divided into three regions which are marked by I, II and III in figure\,\ref{fig:tauSpectDrawing}. 
The high field maxima in regions I and III are used as longitudinal confinement for UCNs in their low-field seeking spin state. 
In between (region II), a low field is required to maintain the neutron polarization during storage.
The superconducting coils are connected in series. Therefore, the general field shape cannot be changed, but the field magnitude can be scaled by varying the current through the coils in the range of \SIrange{0}{100}{\ampere}.
Radial confinement of UCNs is provided by permanent magnets in a Halbach octupole configuration \cite{Halbach1980}. 
It consists of 24 identical rings, each composed of $N =$ 32 wedge-shaped segments with a height of $L=\SI{57.5}{\milli\meter}$ (see figure\,\ref{fig:HB_fig}a). 
All 768 pieces are made from $\isotope[]{Sm}_2\isotope[]{Co}_{17}$ with a nominal remanence of $B_R=\SI{1.1}{\tesla}$ and relative permeability of $\mu_r=1.1$. 
The whole structure is $\SI{1.38}{\meter}$ long, with an inner radius of $R_i=\SI{54.0}{\milli\meter}$ and outermost radius of $R_o=\SI{84.1}{\milli\meter}$. 
It was custom-made by Arnold Magnetic Technologies AG, Lupfig, Switzerland.
Details on the design of the Halbach octupole can be found in \cite{KarchPhD}.
Theoretically, the resulting magnetic flux density inside an infinitely long multipole of polarity $2k$ (for an octupole $k =$ 4) made from $N$ wedge-shaped elements is given in cylindrical coordinates ($r, \theta$) by \cite{Halbach1980,Bluemler2021}\\
\begin{equation}
\label{eq:Halbach_multipole}
\vec{B}(r,\theta) =F_{geom}  \; r^{\:(k-1)} \left[\begin{array}{c}{\cos[(k-1)\theta + k\alpha ]}\\ {-\sin[(k-1)\theta + k\alpha ]}\\\end{array}\right],
\end{equation}
with
\begin{eqnarray}
\label{eq:Halbach_Fgeom}
F_{geom}  =& \frac{k  B_R}{(k-1) \sqrt{\mu_r}}\left(\frac{1}{R_i^{(k-1)}}-\frac{1}{R_o^{(k-1)}}\right)
\cdot\\
&\frac{\sin{((k+1)\pi/N)}}{(k+1)\pi/N} \frac{N\sin{(2\pi/N)}}{2\pi}
\end{eqnarray}
%
The factor $F_{geom}$ is the scaling of the principal field due to the particular geometry and materials. 
The angle $\alpha$ describes the rotation angle of the cylinder around its axis. 
If such an infinitely long structure is truncated to a length $L$, $F_{geom}$ has to be extended to give its central field ($z = L/2$) by the following approximation by integrating magnetic dipoles (for details see \cite{Bluemler2021}):\\
\begin{equation}\label{eq:HB_length}
F'_{geom} =  F_{geom} \frac{\scriptstyle{L(L^8+18L^6R^2+126L^4R^4+420L^2R^6+630R^8)}}{\scriptstyle{(4R^2+L^2)^{9/2}}},
\end{equation}
with $R = \frac{R_o + R_i}{2}$ the central radius.
Figure\,\ref{fig:HB_fig}b shows excellent agreement of this equation with finite elements (FEM) simulations (COMSOL Multiphysics\footnote{https://www.comsol.com/}) and some measurements with Hall-probes. 
The magnetic flux density was measured at a distance of $\SI{2.5}{\milli\meter}$ with respect to the octupole's inner surface and was extrapolated using the $r^3$ dependency in \ref{eq:Halbach_multipole}, resulting in an average magnetic flux density of $B(r=\SI{53.5}{\milli\meter})=\num{0.902\pm0.026}\si{\tesla}$. 
The minimum flux density was found as $B_r^{\text{min}}(r=\SI{53.5}{\milli\meter})=\SI{0.779}{\tesla}$.\\
The process of adiabatic fast passage spin flipping (see section\,\ref{ssec:afpsf}) requires small radial magnetic field components to work efficiently.
%
While the axial field shows sufficient homogeneity, the strong radial component of the octupole certainly needs compensation.
This can be achieved by using a second octupole, coaxially aligned and rotated by an angle $\alpha = \pi/k = \SI{45}{\degree}$ to the first.
If its geometry is chosen such that $F'_{geom}$ matches that of the other octupole, 
the magnetic fields inside cancel each other (cf. \ref{eq:Halbach_multipole}) \cite{Leupold1993,Bluemler2021}.
Therefore, such a second, smaller octupole for compensation of the larger octupole for storage (see figure\,\ref{fig:HB_fig}a) 
 is mounted together with the RF-coils of the adiabatic fast passage (AFP) spin flipper \cite{Abragam1961} so that both can be moved in and out of the trapping region (see section\,\ref{ssec:afpsf}). 
To guarantee identical behavior when the experiment is cooled down, the second octupole was made from the same material with $N =$ 32, $R_i=\SI{41.2}{\milli\meter}$, $R_o=\SI{47.0}{\milli\meter}$, 
 and $L=\SI{150}{\milli\meter}$. 
These dimensions were optimized using FEM-simulations, however equation~\ref{eq:HB_length} also gives $ F'_{geom}=\SI{6226}{\tesla/\cubic\meter}$ 
compared to $F'_{geom}=\SI{6032}{\tesla/\cubic\meter}$ for the storage octupole (cf. figure\,\ref{fig:HB_fig}b). 
The individual segments of the octupole magnet were glued into a holding shell made from titanium ($V_F=\SI{-48}{\nano\electronvolt}$). 
This material was chosen because it is non-magnetic and allows for a small wall thickness while maintaining structural stability. \\
In order to quantify the compensation, the radial field of the storage octupole was measured at room temperature before the installation in \tss with and without the compensation octupole \cite{RossPhd}. The fields and gradients were measured at a radius of $r=\SI{25}{\milli\meter}$, corresponding to the outside surface of the neutron guide in the compensation octupole.
At this radius, the average magnetic field and gradient were reduced from $B=\SI{75\pm5}{\milli\tesla}$ and $|\nabla_rB|=\SI{112\pm54}{\milli\tesla/\centi\meter}$ to $B=\SI{2.2\pm0.8}{\milli\tesla}$ and $|\nabla_rB|=\SI{2.1\pm0.8}{\milli\tesla/\centi\meter}$, which is sufficiently low to allow for the spin flip efficiency to reach $\epsilon_{\rm{SF}} \sim 1$.
It must be taken into account that axial misalignment leads to incomplete compensation of the magnetic field.
The alignment of the compensation octupole leading to the described compensation was performed with an axial offset of only $\delta y =\SI{0.2}{\milli\meter}$.\\
The superposition of the longitudinal and the radial field creates a central region which is surrounded by increasing magnetic flux density, as schematically shown in figure\,\ref{fig:3dFields}.
While the longitudinal field can be changed and thus the trap potential can be varied, this is not possible for the radial field being made from permanent magnets. 
In order to maximize the trap volume, the optimum current in the $B_z$ producing superconducting coils was determined: 
On the one hand, both field maxima (regions I and III) should be larger than the minimum flux density of the octupole $B_r^{\text{min}}(r=\SI{53.5}{\milli\meter})$ so that the trap depth is limited by the fixed octupole field. 
On the other hand, the longitudinal holding field in region II must be low, as otherwise the storable phase space decreases (Sec.\,\ref{ssec:afpsf}). 
The optimum current was determined as $I=\SI{33}{\ampere}$, leading to $B_z^{\text{max}}=\SI{1.023}{\tesla}$ in region I and $B_z^{\text{max}}=\SI{2.050}{\tesla}$ in region III, as well as a holding field between $B_z=\SI{202}{\milli\tesla}$ and $\SI{208}{\milli\tesla}$ in region II.
The magnetic trap potential is then given, as $V_{\text{magn}}=-\mu_n B_r^{\text{min}}(r=\SI{53.5}{\milli\meter}) = \SI{47.0}{\nano\electronvolt}$ with $\mu_n = \SI{-60.3}{\nano\electronvolt/\tesla}$ the neutron magnetic moment.
The trap has a volume of up to $\sim \SI{10.7}{\liter}$ for UCNs at maximum energies. Neutrons with lower energies occupy a smaller volume.
The compensation octupole and spin flipper form the SFU, which along with the neutron guide is pulled out of the trapping region during storage by a translation mechanics operated by a stepper motor. 
The range of motion is between $z=\SI{397}{\milli\meter}$ and $\SI{1200}{\milli\meter}$ (see figure\,\ref{fig:sf_and_det_positions}). 
Spring-mounted wheels are attached to the front of the compensation octupole at $\SI{45}{\degree}$ relative to the vertical axis to reduce friction on the storage octupole surface. 
A copper shield ($V_F = \SI{168}{\nano\electronvolt}$) on the front face reflects neutrons during the filling process as long as the SFU is still inside the trap. 
A dedicated holding construction prevents the compensation octupole from rotation around the $z$-axis \cite{KahlenbergPhD}.
%
\subsection{Adiabatic fast passage spin flipper}
\label{ssec:afpsf}
\begin{figure}[b]
\centering
  \includegraphics[width=0.99\columnwidth]{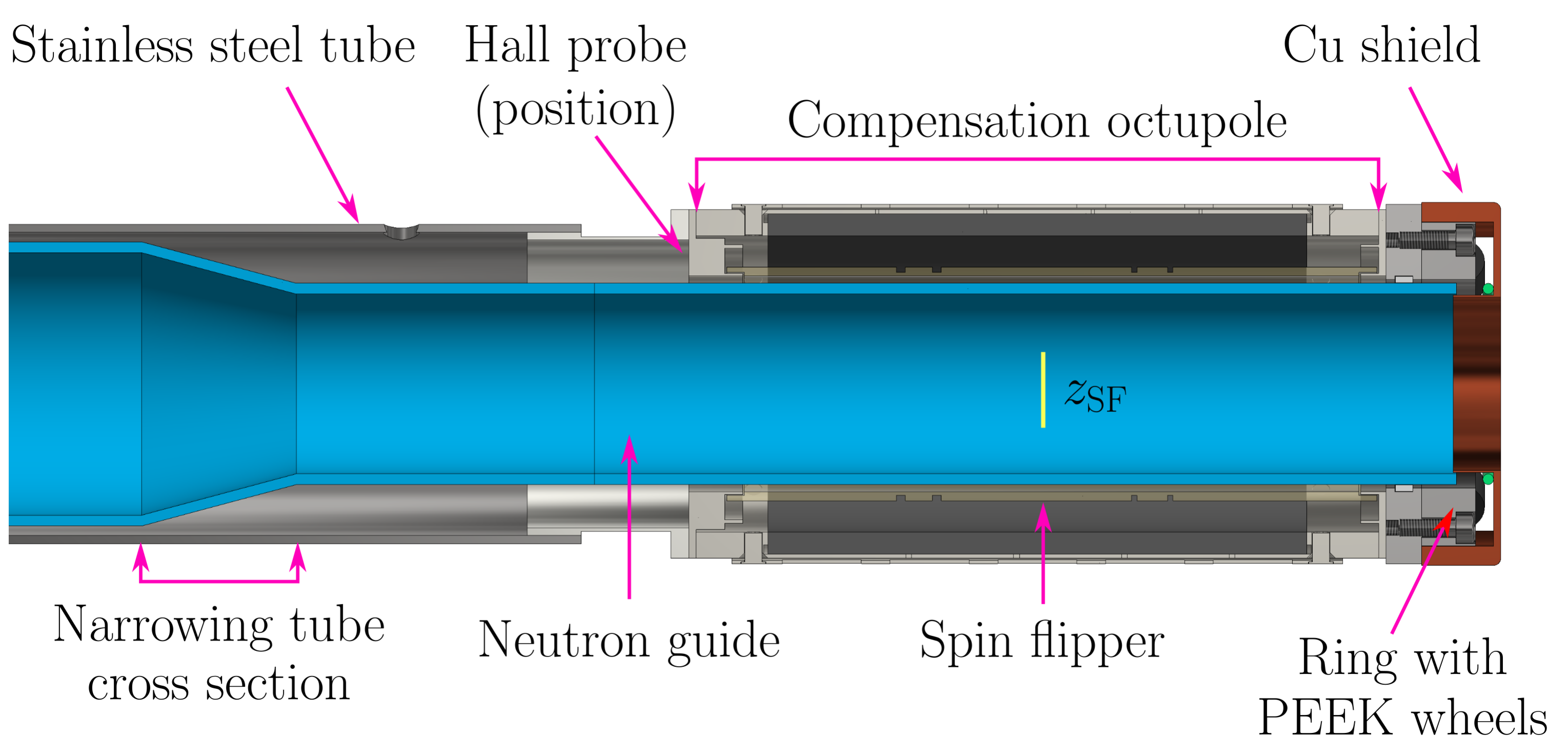}
\caption{The SFU as CAD rendering. The spin flipper with its center position marked with $z_{SF}$ is surrounded by the compensation octupole. A ring with wheels made from PEEK reduces friction when moving the SFU on the storage octupole surface. A copper shield ($V_F=\SI{168}{\nano\electronvolt}$) is fixed at the front face of the SFU to back reflect neutrons while the SFU is at filling position. Due to spatial constraints, the neutron guide's inner radius is reduced from $r=\SI{36.5}{\milli\meter}$ to $r=\SI{25.0}{\milli\meter}$. The stainless steel tube provides mechanical support and is connected to the translation stage (see figure\,\ref{fig:tauSpectDrawing}). A uniaxial Hall probe is used to measure the longitudinal component of the magnetic field.}
\label{fig:sfu}
\end{figure}

\begin{figure}[t]
\centering
 \includegraphics[width=0.95\columnwidth]{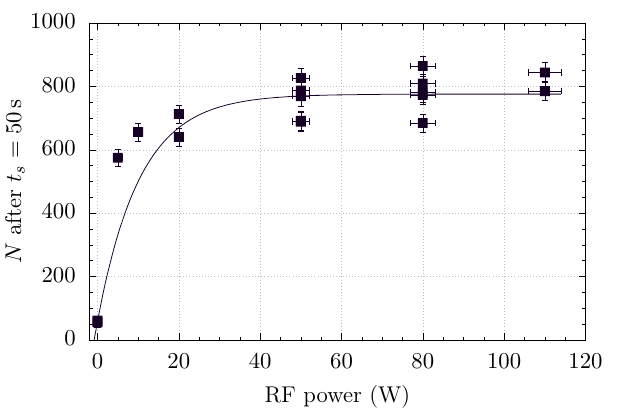}
\caption{Number of stored neutrons depending on the irradiated RF power at $\zsffill = \SI{980}{\milli\meter}$ in SF2. The data show a saturation behavior of the neutron counts beyond an RF power of $\SI{50}{\watt}$. The line is an exponential function (\ref{eq:exponential_function}) plotted to guide the eye.}
\label{fig:n_vs_power}
\end{figure}

Ultracold neutrons  enter \tss through stainless steel tubes (inner diameter $d_i = \SI{66}{\milli\meter}$). Inside the cryostat's vacuum a telescopic neutron guide system, consisting of a fixed inner stainless steel tube and a movable outer fused silica tube ($V_F = \SI{95}{\nano\electronvolt}$ and inner diameter $d_i = \SI{73}{\milli\meter}$) is used to guide the neutrons into the storage region.
Ultracold neutrons  enter \tss through a telescopic neutron guide system consisting of stainless steel tubes (inner diameter $d_i = \SI{66}{\milli\meter}$, $V_F = \SI{180}{\nano\electronvolt}$) outside and a fused silica tube within the cryostat ($d_i = \SI{73}{\milli\meter}$).
An adiabatic fast passage (AFP) spin flipper \cite{Abragam1961} is located at the end of the fused silica tube.
To accommodate the spin flipper, the tube diameter is reduced to $d_i = \SI{50}{\milli\meter}$ (see figures\,\ref{fig:tauSpectDrawing} and \ref{fig:sfu}).
As soon as the UCNs enter the high magnetic field in region I (see figure\,\ref{fig:tauSpectDrawing}) they are separated into \textit{high-field seekers} (HFS) and \textit{low-field seekers} (LFS).
Only HFS (as well as LFS with kinetic energies larger than the magnetic potential in region I) are able to pass the field maximum. 
As high-field seeking neutrons are not storable, the AFP spin flipper is used to rotate the neutron spin by $\SI{180}{\degree}$ and thus to transform HFS into storable LFS\footnote{The fraction of LFS that pass the field maximum in region I are spin flipped to HFS which are not storable and leave the trap.}. 
An AFP spin flipper for neutrons has been proposed by  V. Luschikov \cite{Luschikov1984} and several have been successfully employed in other experiments \cite{Geltenbort2009,Holley2012}. Its application in the context of a vertical magneto-gravitational trap for a neutron lifetime experiment is also discussed in \cite{Zimmer2000}. 
\begin{figure}[b]
  \includegraphics[width=0.99\columnwidth]{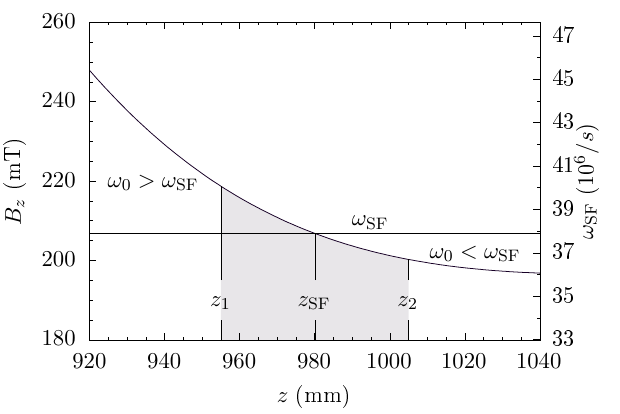}
\caption{Longitudinal magnetic field in the spin flipping region around the nominal spin flip position, $\zsffill$ with the field magnitude on the left axis and the required spin flipper angular frequency $\omegasf$ calculated using \ref{eq:sf_frequency}. The gray box marks the physical extent of the spin flipper between $z_1$ and $z_2$.}
\label{fig:AFP_gradient}
\end{figure}
\begin{figure*}[t]
\centering
  \includegraphics[width=0.9\textwidth]{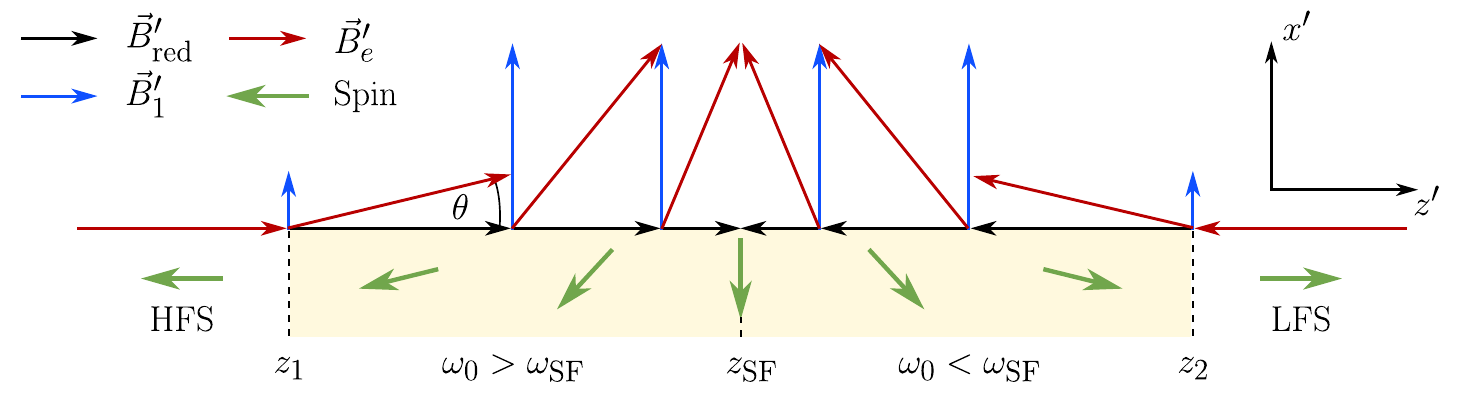}
\caption{Sketch of the spin flip of an initial HFS UCN. From the left, the UCN enters the spin flipper (yellow) with its spin (green) antiparallel aligned to $\vec{B}'_e(z)$. $\vec{B}'_1$ points along the $\hat{x}'$ axis and tilts $\vec{B}'_{\text{red}}(z)$ by an angle $\theta$. On the left side of the resonance at $\zsffill$, $\vec{B}'_{\text{red}}(z)$ points along $z'$ while on the right side it points to the opposite direction. The neutron spin follows $\vec{B}'_e(z)$ adiabatically and the UCN exits the spin flipper as LFS. Arrow lengths are not to scale and for illustrative purposes only.}
\label{fig:AFP}
\end{figure*}
The principle of operation is based on nuclear magnetic resonance and visualized in figure\,\ref{fig:AFP}.
In a magnetic field $\vec{B}$, the neutron spin precesses at the Larmor frequency $\omega_0 = -\gamma |\vec{B}|$ with the gyromagnetic ratio of the neutron $\gamma=\num{-1.832\,471\,71(43)}\times\SI{e8}{\per\tesla\per\second}$ \cite{CODATA2018}.
In the spin flipping region around $z = \SI{980}{\milli\meter}$, the magnetic field has a predominant gradient in the longitudinal direction, while radial magnetic field components caused by the storage octupole are suppressed using the compensation octupole magnet. Therefore, the Larmor frequency is position dependent as shown in figure\,\ref{fig:AFP_gradient}. 
In order to perform a spin flip, a perpendicular ($xy$-plane) rotating  magnetic field $B_1$ is required.
With the center of the spin flipper being located at a nominal position $\zsffill$ ('resonance position'), this resonance frequency is given by 
\begin{equation}\label{eq:sf_frequency}
 \omegasf = \omega_0 = -\gamma |\vec{B}(\zsffill)|.
\end{equation}
In a reference frame which rotates at the spin flipper frequency (indicated by primes) $\omegasf$, the spin precesses at the so-called reduced frequency $\omega_{\text{red}}' = \omega_0 - \omegasf$ around the locally reduced field
\begin{equation}
 \vec{B}_\text{red}'(z) = \left(\frac{-\omega_{\text{red}}'(z)}{\gamma}\right) \hat{z}' = \vec{B}_\text{z}(z) +  \left( \frac{\omegasf}{\gamma} \right) \hat{z}' .
\end{equation}
In this reference frame, $\hat{z}=\hat{z}'$ and $\vec{B}_1'$ is static and points along an arbitrarily chosen axis $\hat{x}'$.
The superposition of $\vec{B}_1'$ and $\vec{B}_\text{red}'(z)$ gives the effective field the neutrons experience in the rotating reference frame
\begin{equation}
\vec{B}_{e}'(z) = \vec{B}_\text{red}'(z) + \vec{B}_1'  = \left( B_z(z) + \frac{\omegasf}{\gamma}\right)\hat{z}'+ B_1\hat{x}'.
\end{equation}
The angle between this effective field and the $\hat{z}'$-axis is given by
\begin{equation}\label{eq:theta_angle}
 \tan{(\theta)} = \frac{B_1'}{\raisebox{-0.8mm}{$B_z(z) + \frac{\omegasf}{\gamma}$}} = \frac{\omega_1}{\raisebox{-0.8mm}{$\omega_\text{red}'(z)$}}.
\end{equation}
Due to the negative field gradient of $B_z$, $\omega_{\text{red}}>0$ for $z<\zsffill$ and $\omega_{\text{red}}<0$ for $z>\zsffill$.
This results in a change of sign in $\theta$ and thus in the orientation of $\vec{B}_{e}'$.
For example, the spin of an HFS neutron follows this field adiabatically thus changing its orientation as shown in figure\,\ref{fig:AFP} and the neutron exits the spin flipper as LFS.
\begin{figure}[b]
\centering
  \includegraphics[width=0.99\columnwidth]{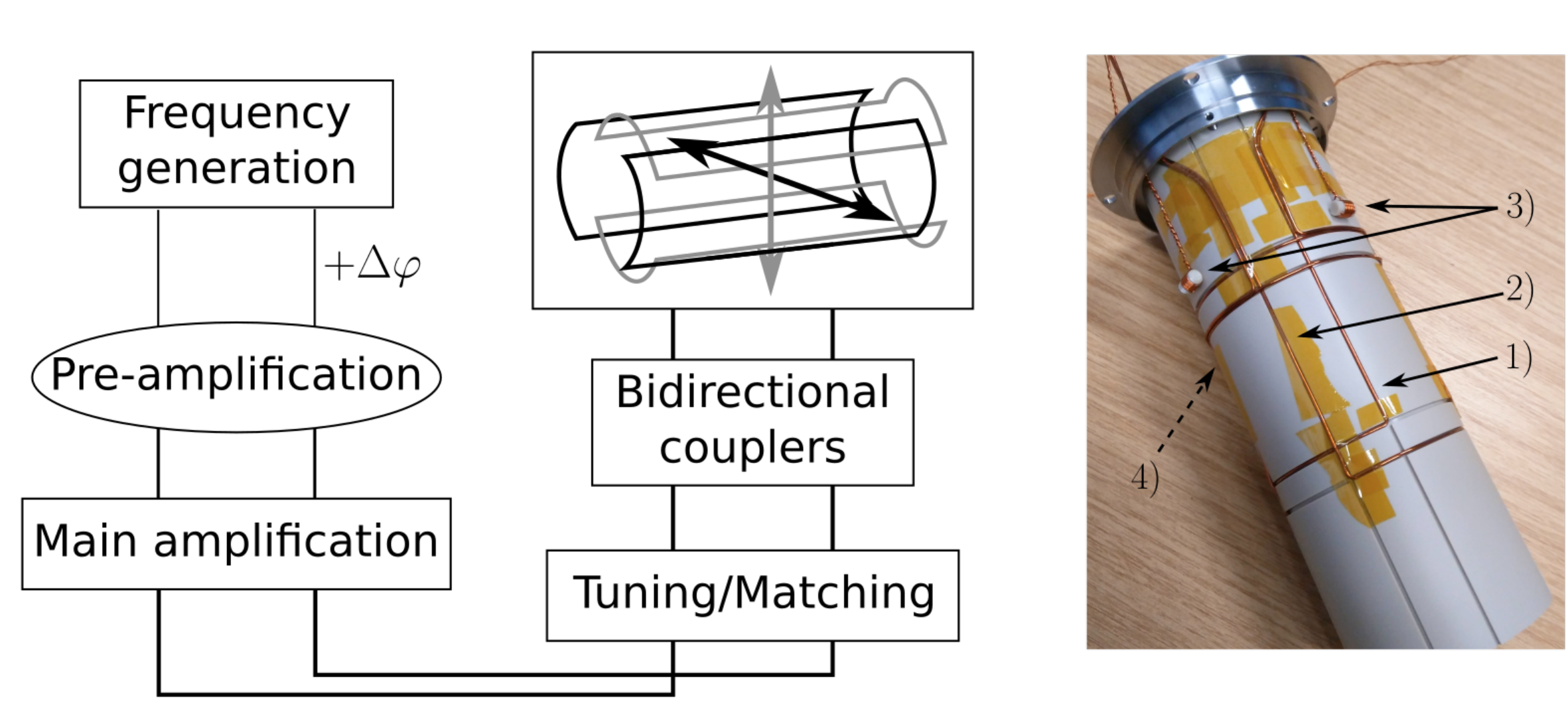}
\caption{Schematic of the $B_1$ field generation circuit (left) and the finished spin flipper SF2 (right). The spin flipper consists of two saddle coils (1) and (2) which are wound on a holding cylinder made from PEEK. Two pick-up coils (3) and a temperature sensor (4) (not visible) can be used for functionality and reproducibility checks.}
\label{fig:sc_circuit}
\end{figure}
The adiabatic fast passage spin flipper used for the \tss experiment consists of two orthogonal saddle coils wound on a holding Polyether-ether-ketone (PEEK) cylinder (see figure\,\ref{fig:sc_circuit}).
The coils have an opening angle $\phi = \SI{120}{\degree}$, half diameter $d/2=\SI{32.5}{\milli\meter}$ and a length of $l=\SI{60}{\milli\meter}$.
A longitudinal offset of $\SI{10}{\milli\meter}$ between the coils is added to reduce cross-talk, decreasing the spin flipper to an effective length of $l_{\text{eff}} = \SI{50}{\milli\meter}$. 
An individual RF signal is fed into each coil, resulting in a linearly polarized magnetic field (see figure\,\ref{fig:sc_circuit}).
By adjusting the phase between the two signals, the overall polarization of the $B_1$ field can be set to circular ($\pm \sigma$), linear ($\pi$) or any other elliptic polarization.
The RF excitation signals are generated by the two outputs of a frequency generator \footnote{Siglent SDG2042X} and amplified in two steps \footnote{First stage: K-N-E power amplifier 10W-QRP, second stage: RM Italy HLA300 V plus}.
Frequency tuning and impedance matching of the excitation circuits to $Z=\SI{50}{\ohm}$ is done by radio tuners \footnote{MFJ-969 Versa Tuner II} outside the closed cryostat.
Bidirectional couplers within the signal lines are used to monitor and subsequently adjust the phase between the excitation signals.
A second very similar spin flipper has been built to implement the double-spin flip loading scheme described in section\,\ref{sec:phase_space}.
The spin flippers are named according to the order of passage by the UCNs from the experiment entrance. The SF1 is located at a lower z-position than SF2.
Figure\,\ref{fig:n_vs_power} shows the spin flip efficiency of SF2 as a function of the irradiated RF power.
$\SI{50}{\second}$ storage measurements were used at the spin flip position $z=\SI{980}{\milli\meter}$, without the spectral cleaning procedure. 
The horizontal axis shows the RF power ($P$) which is related to the amplitude $B_1$, as $B_1\propto \sqrt{P}$.
The line is an exponential function
\begin{equation}\label{eq:exponential_function}
 N = a\left( 1-\exp\left(-\frac{P}{b}\right)\right)+c
\end{equation}
with the number of neutrons after a storage time of $\SI{50}{\second}$ and the fit parameters $a$, $b$ and $c$ for scaling and consideration of an offset.
We find that the neutron density in the trap saturates from $P\sim \SI{50}{\watt}$ onward, indicating a maximum spin flip efficiency\footnote{The maximum spin flip efficiency does not automatically mean $\varepsilon_{\text{SF}}=1$ as the absolute spin flip efficiency could not be measured.} from there on.
The irradiated RF power in all measurements with \tss is between $P=\SI{80}{\watt}$ and $\SI{100}{\watt}$.
The RF phase $\Delta \varphi$ between the excitation signals of the saddle coils was scanned around the value of $|\Delta \varphi| = \SI{90}{\degree}$ confirming the expected phase setting \cite{RossPhd}.
\subsubsection{Single spin flip energy acceptance}
\label{sec:filling}
In order to maximize the UCN density per fill, the energy acceptance of the trap should be as large as possible.
\begin{figure}[b]
\centering
 \includegraphics[width=0.95\columnwidth]{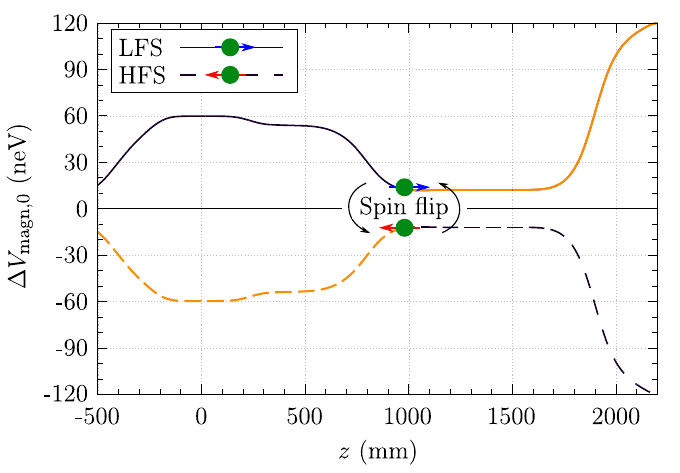}
\caption{Magnetic potential as experienced by LFS (solid line) and HFS (dashed line). Due to energy conservation, HFS (LFS) are accelerated (decelerated) when they enter an increasing field and decelerated (accelerated) in a decreasing field. The spin flip changes the sign of $\Delta V_{\text{magn},0}$. The potential as seen from an initial HFS spin flipped to an LFS is marked in orange.}
\label{fig:sSF_potentials}
\end{figure}
We consider a neutron in the high-field seeking state ($\vec{\mu}_n \uparrow\uparrow \vec{B}$) with an initial kinetic energy $E_{\text{kin},0}$, see figure\,\ref{fig:sSF_potentials}.
Before entering \tss, the neutron's potential energy due to the magnetic field is $V_{\text{magn},0}=\SI{0}{\nano\electronvolt}$.
While moving into the high field region, the HFS neutron gains the kinetic energy $\Delta E_{\text{kin}} = |\mu_n|B_{z}^{\text{max,I}} = \SI{61.7}{\nano\electronvolt}$.
After the field maximum, it is decelerated again until it reaches the position of the spin flipper $\zsffill$. 
At that point, the HFS neutron has gained kinetic energy of $\Delta E_{\text{kin}}(\zsffill) = -\Delta V_{\text{mag},0}(\zsffill)=|\mu_n|B_z(\zsffill)$.
After the spin flip, the magnetic moment is aligned antiparallel with $B_z$ (LFS) and $\Delta V_{\text{magn},0}$ becomes positive.
The total energy of the now LFS neutron is thus given by

\begin{eqnarray}\label{eq:energy_no_grav}
E(\zsffill) &=& E_{\text{kin},0} + \Delta E_{\text{kin}}(\zsffill) + \Delta V_{\text{magn},0}(\zsffill) \\
 &=& E_{\text{kin},0} + 2|\mu_n| B_z(\zsffill)
\end{eqnarray}
If its total energy is below the trap potential (see Fig.~\ref{fig:3dFields}), $V_{\text{magn}} = \SI{47.0}{\nano\electronvolt}$  an LFS neutron is storable. 
Therefore, the maximum allowed initial kinetic energy is
\begin{equation}\label{eq:energy_acceptance_no_gravity}
 E_{\text{kin},0}^{\text{max}}(\zsffill) = \SI{47.0}{\nano\electronvolt}-2|\mu_n|B_z(\zsffill).
\end{equation}
So far, we ignored gravity.
To account for gravity, we need to consider the height difference $s$ between the position of the spin flip and the lowest point of the storage octupole surface.
The concentrically aligned neutron guide and storage octupole have different radii ($r_{\text{guide}}=\SI{25}{\milli\meter}$ compared to $r_{\text{SO}} = \SI{54}{\milli\meter}$).
Assuming the spin flip can take place at any height within the neutron guide, $s$ can take values from \SI{29}{\milli\meter} to \SI{79}{\milli\meter} given by the difference and sum of the radii.
The maximum allowed initial kinetic energy in \ref{eq:energy_acceptance_no_gravity} is reduced by $\Delta V_{\text{grav}}(s)=s\cdot \SI{102.4}{\nano\electronvolt/\meter}$ and results in
\begin{equation}\label{eq:total_energy_storage_condition}
E_{\text{kin},0}^{\text{max}}(\zsffill, s) = \SI{47.0}{\nano\electronvolt}-2|\mu_n|B_z(\zsffill) - \Delta V_{\text{grav}}(s).
\end{equation}
For a nominal spin flip position $\zsffill=\SI{980}{\milli\meter}$ of 
\begin{equation}
    E_{\text{kin},0}^{\text{max}}(\SI{980}{\milli\meter}, \SI{29}{\milli\meter} ) = \SI{18.9}{\nano\electronvolt}.
\end{equation}
So far, any correlation of the UCN energy with their vertical position in UCN guide (energy staggering) is not taken into account.
%
\subsubsection{Double spin flip energy acceptance}
\label{sec:phase_space}
\begin{figure}[b]
\centering
 \includegraphics[width=0.95\columnwidth]{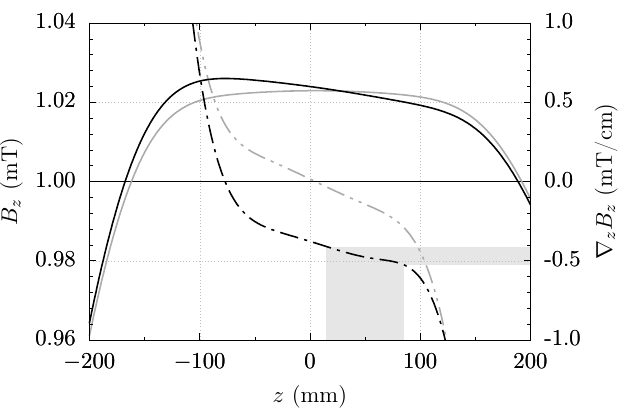}
\caption{Longitudinal magnetic field $B_z$ around the maximum of the field in region I without (gray) and with the gradient coils C3 and C5 (black). Solid lines refer to the magnetic field (left axis), dashed lines to the field gradient (right axis). The shaded area shows the longitudinal range and the field gradient covered by SF1. The gradient at $\zsffillfirst=\SI{50}{\milli\meter}$ is $\nabla_z B_z(\zsffillfirst) = \SI{-0.48}{\milli\tesla/\centi\meter}$ for  $I_{\text{C3}} = \SI{20}{\ampere}$ and $I_{\text{C5}} = \SI{-10}{\ampere}$.}
\label{fig:gradient_SF1}
\end{figure}

\begin{figure}[t]
\centering
 \includegraphics[width=0.95\columnwidth]{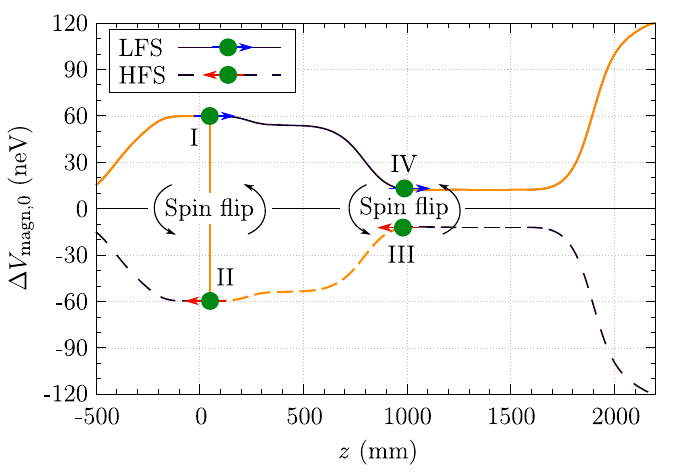}
\caption{Magnetic potential as experienced by LFS (solid line) and HFS (dashed line). The potential energy of an initial LFS during the double spin flip is shown by the orange line. It is decelerated as the potential energy rises. At SF1, it is converted to a HFS with $\Delta V_{\text{magn},0}$ changing sign (from I to II). The HFS is decelerated again until at SF2 $\Delta V_{\text{magn},0}$ changes sign again (from III to IV). 
The neutron kinetic energy is reduced by $\SI{110.5}{\nano\electronvolt}$ at $\zsffillsecond=\SI{980}{\milli\meter}$. The potential energy at SF2 is $\SI{12.5}{\nano\electronvolt}$.}
\label{fig:sSF_dSF_potentials}
\end{figure}

In order to increase the energy acceptance of \tss, an upgrade of the experiment introduced a double AFP spin flip technique using a second spin flipper (SF1) in the high field of region I. 
For the AFP spin flipper to work we impose a longitudinal field gradient of $\nabla_z B_z(\zsffillfirst) = \SI{-0.48}{\milli\tesla/\centi\meter}$ using coils, named C3 and C5, inside the cryostat in anti-Helmholtz configuration (see figure\,\ref{fig:gradient_SF1}).
Using this technique, LFS neutrons are at first decelerated in the increasing $B_z$ field  by $|\mu_n|B_z(\zsffillfirst) = \SI{61.5}{\nano\electronvolt}$ (see figure\,\ref{fig:sSF_dSF_potentials}), then the neutron spin state is flipped from LFS to HFS and the neutrons are subsequently decelerated by $|\mu_n|B_z(\zsffillfirst)-|\mu_n|B_z(\zsffillsecond) = \SI{49.0}{\nano\electronvolt}$ on their way to the spin flipper in the low field region (SF2) ($|\mu_n|B_z(\zsffillsecond)=\SI{12.5}{\nano\electronvolt}$).
Spin flipper 2 converts HFS to storable LFS neutrons (Pos. III to IV). 
The neutron energy after passage of SF2 is given by
\begin{equation}
 E = \left(E_{\text{kin},0}-\SI{110.5}{\nano\electronvolt}\right) + V_{\text{magn}}(\zsffillsecond) + \Delta V_{\text{grav}}( s).
\end{equation}
With the trapping condition $E\leq\SI{47.0}{\nano\electronvolt}$  follows the maximum allowed energy $E_{\text{kin},0}^{\text{max}}(\zsffillsecond)=\SI{142.0}{\nano\electronvolt}$ for an incoming neutron to be stored in the trap potential.
Due to the deceleration on their way into the trap, neutrons will be reflected if their initial kinetic energy is less than $E_{\text{kin},0}^{\text{min}}(\zsffillsecond)=\SI{110.5}{\nano\electronvolt}$.
Therefore, the accepted initial energy interval using the  double spin flip technique is $[\SI{110.5}{\nano\electronvolt},\SI{142.0}{\nano\electronvolt}]$.
This is an increase in interval width by \SI{66}{\percent} compared to the single spin flip case.
Combined with the $\sqrt{E}$-dependency of the neutron density per unit energy $\varrho(E)$ at the exit of the source \cite{Golub1991} this results in a higher number of neutrons stored.
Since the neutrons are decelerated along $B_z$, only neutrons with a large enough longitudinal velocity will reach the trap.
%
To account for this, the minimum kinetic energy can be rewritten as 
\begin{equation}
    \label{eq:min_kin_energy_angle}
    E_{\text{kin},0}^{\text{min}}(\zsffillsecond)=\SI{110.5}{\nano\electronvolt}/\cos^2(\theta),
\end{equation}
where $\theta$ is the angle between the velocity vector and the longitudinal field.
For an angle $\theta = \SI{28.1}{\degree}$, the minimum energy reaches $E_{\text{kin},0}^{\text{min}} = \SI{142.0}{\nano\electronvolt}$ and the kinematic acceptance becomes zero.
A comparison of both spin flip techniques has been done using Monte Carlo simulations (see Sec.\,\ref{ssec:sim_dsf_technique}) 
Since the two spin flip techniques work for different energy and therefore velocity ranges, a combination of both techniques can lead to a higher number of neutrons in the trap than one technique alone.
This mixed filling technique is currently under investigation.
%
\subsection{In-situ UCN detector}
\label{ssec:detector}
\begin{figure}[b]
\centering
  \includegraphics[width=0.99\columnwidth]{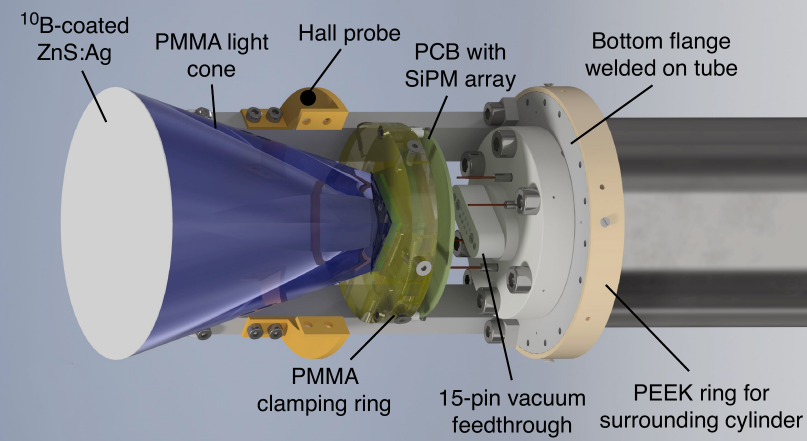}
\caption{CAD rendering of the UCN detector. For details, see main text. }
\label{fig:detSketch}
\end{figure}
The design of the magnetic trap and the environmental conditions within the cryostat impose several constraints on the UCN detector. 
\begin{enumerate}
        \item It must be suitable for high vacuum, cryogenic temperatures and high magnetic fields of up to $\SI{2}{\tesla}$.
        \item It has to be possible to move the detector inside $\tau$SPECT to different positions.
        \item It must be able to detect UCNs efficiently under these conditions.
\end{enumerate}
The construction of our custom UCN detector and its motion mechanism is described in \cite{KahlenbergPhD}, the schematic setup is shown in figure\,\ref{fig:detSketch}. 
UCNs are captured in an $\SI{80}{\nano\meter}$ thick $\isotope[10]{B}$ layer coated\footnote{Coating done by CDT CASCADE Detector Technologies GmbH, \url{https://n-cdt.com/}.} on a silver-doped zinc sulphide (ZnS:Ag) scintillator of \SI{22.6}{\micro\meter} thickness. 
The $\isotope[10]{B}$/$\isotope[11]{B}$ ratio\footnote{$\isotope[10]{B}$ enrichment is determined by the base material used for coating and is guaranteed by CDT CASCADE Detector Technologies GmbH at $>96\%$.} was chosen to obtain $V_F \approx \SI{0}{\nano\electronvolt}$.
The charged particles from the nuclear $\isotope[10]{B}(n,\alpha)\isotope[7]{Li}$ reaction generate light in the scintillator, which is guided towards a $4\times 4$ silicon photomultiplier (SiPM) array (Hamamatsu S13361-6050AE-04) by a cone-shaped light guide made from Poly-methyl-methacrylate (PMMA). 
The 16 channels are summed and amplified in two steps inside and outside the experiment, and are subsequently digitized by a 14-bit ADC with a sampling rate of $\SI{10}{Msps}$. 
\begin{figure}[b]
\centering
  \includegraphics[width=0.99\columnwidth]{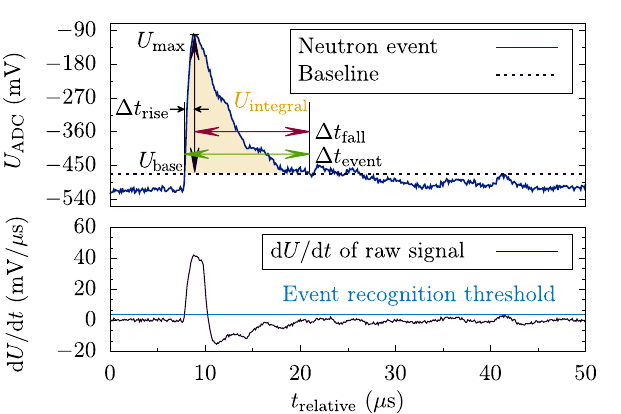}
\caption{Single event in the detector raw data (top) and the time derivative of the same signal (bottom). If the time derivative exceeds the event recognition threshold, the event is analyzed by the pattern recognition algorithm with respect to the event maximum voltage $U_{\text{max}}$, the integrated voltage $U_{\text{integral}}$, the rise time $\Delta t_{\text{rise}}$, the fall time $\Delta t_{\text{fall}}$ and the overall event length $\Delta t_{\text{event}}=\Delta t_{\text{rise}}+\Delta t_{\text{fall}}$. The baseline voltage is defined as the voltage $U_{\text{base}}$ when the event recognition threshold was exceeded.}
\label{fig:neutron_event}
\end{figure}
Raw data is recorded continuously during each measurement run (see Sec.\,\ref{sec:procedures} for the full sequence of each run), an offline pattern recognition algorithm subsequently detects neutron events in the following way:
A sharply rising edge triggers a new event.
Parameters like rise- and fall-times, peak voltage with respect to baseline voltage, event length and voltage integral over the event are evaluated and compared to empirical threshold values. 
An example event is shown in figure\,\ref{fig:neutron_event}, the algorithm is described in detail in \cite{RossPhd}.
The performance of the custom UCN detector for \tss was compared to a commercial CASCADE-U detector\footnote{UCN detector by \href{cdt CASCADE Detector Technologies GmbH}{https://n-cdt.com}} at room temperature: 
UCNs were stored for $t_s = \SI{20}{\second}$ in an aluminum storage bottle\footnote{Using a storage bottle with aluminum walls leads to a very soft energy spectrum of stored neutrons, similar to the expected energy spectrum in \tss .} ($V_F = \SI{54}{\nano\electronvolt}$, $V=\SI{9.67\pm0.02}{\liter}$ \cite{KahlenbergPhD}) and counted afterward.  
The relative detection efficiency can then be expressed in terms of UCN counts $N$ and results in
\begin{equation}
 \frac{N_{\text{\tss}}}{N_{\text{CASCADE}}}=1.06(7).
\end{equation}
In \tss, the detector is mounted on a translation stage to store the detector outside or to move it inside the storage region to count UCNs.
\subsection{Environmental monitoring}
\label{ssec:monitoring}
The environmental conditions in \tss are permanently monitored.
Two full range pressure sensors (positions marked in figure\,\ref{fig:tauSpectDrawing}), one cold-cathode sensor and one Pirani sensor, read the pressure in \tss.
The temperature is measured by various sensors, which is especially important for the operation of the superconducting coils. 
The temperature inside the storage octupole is approximately $\SI{170}{\kelvin}$.
The detector and the SFU are both equipped with a 1-axis Hall sensor, which can be used to measure the $z$-component of the magnetic field for reproducibility purposes. 
A nuclear magnetic resonance probe is mounted close to SF1.
Two vibration sensors are attached to the outside of the cryostat which can be used for future systematic studies on microphonic heating.
To adjust the height of the experiment, the iron yoke is placed on four heavy duty lifting columns. Distance sensors monitor the height of the experimental setup.
\section{Simulation of the trap filling process}
\label{sec:filling_sim}
The AFP spin flipping process is well understood in regions with small transverse magnetic field gradients \cite{Holley2012}. 
However, the additional transversal field of the compensation octupole (if not fully compensated) makes an analytic description of the spin flip difficult.
Therefore, the filling process was simulated using Monte Carlo methods.  The time evolution of the spin was explicitly considered through numerical integration\footnote{Python3 numerical integrator odeint (\url{https://docs.scipy.org/doc/scipy/reference/generated/scipy.integrate.odeint.html}) from the SciPy-library \cite{2020SciPy}} of Bloch's equation \cite{Bloch1946} 
\begin{equation}\label{eq:Bloch_equation}
 \frac{d\vec{S}}{dt} = \gamma\left(\vec{S}\times \vec{B}\right)
\end{equation}
here $\vec{S}$ is the spin of a neutron in a magnetic field $\vec{B}$ and $\gamma$ is its gyromagnetic ratio.
The magnetic fields of the superconducting solenoids, the storage octupole and the compensation octupole and the two spin flippers were implemented as analytical approximations (see \cite{RossPhd}). 
The kinetic variables were simulated, taking gravity and the magnetic field gradient (Stern-Gerlach) forces into account.

\subsection{Single spin flipper efficiency}
\label{ssec:sim_transversal_fields}
\begin{figure}[t]
\centering
 \includegraphics[width=0.95\columnwidth]{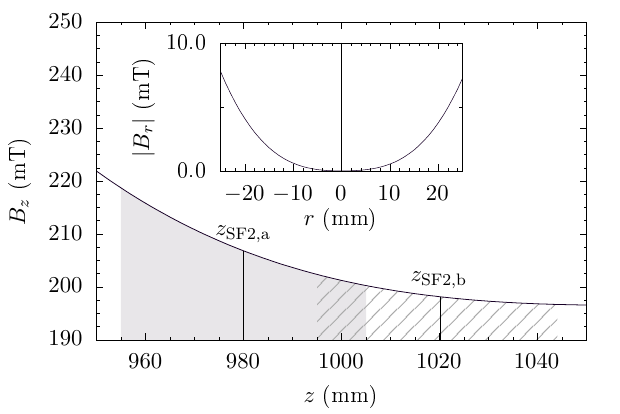}
\caption{Analytic magnetic field approximations used for the Monte Carlo simulation. The inset shows the compensated octupole field on a horizontal line ($y=\SI{0}{\milli\meter}$). Two spin flip positions were simulated, $z_{\text{SF2},\text{a}} = \SI{980}{\milli\meter}$ and $z_{\text{SF2},\text{b}}=\SI{1020}{\milli\meter}$. The solid and dashed areas mark the range covered by the spin flipper in each case.}
\label{fig:sim_fields}
\end{figure}

\begin{figure}[hb]
\centering
 \includegraphics[width=0.95\columnwidth]{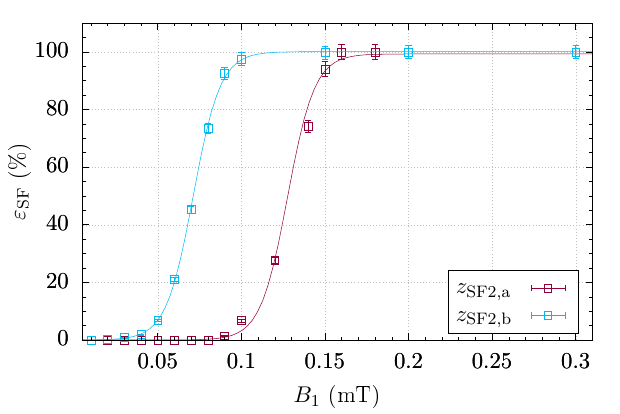}
\caption{Spin flip efficiency $\varepsilon_{\text{SF}}$ depending on the irradiated $B_1$ amplitude as inferred from a Monte Carlo simulation at $z_{\text{SF2},\text{a}}=\SI{980}{\milli\meter}$ and $z_{\text{SF2},\text{b}}=\SI{1020}{\milli\meter}$. The efficiency was calculated using \ref{eq:sf_efficiency}, the lines are logistics functions (\ref{eq:logistics_function}) plotted to guide the eye.}
\label{fig:sim_efficiency_vs_B1}
\end{figure}

The influence of the residual transversal fields was investigated by estimating the efficiency $\varepsilon_{\text{SF}}$ of spin flipping initial HFS neutrons to the LFS state with SF2. 
The simulation was performed for two different spin flip positions $z_{\text{SF2},\text{a}}=\SI{980}{\milli\meter}$ and $z_{\text{SF2},\text{b}}=\SI{1020}{\milli\meter}$ (see figure\,\ref{fig:sim_fields}).
Each simulation comprised a set of the same $N = 4000$ neutrons with fixed but random parameters: starting positions in the $x-y$ plane within the narrowed neutron guide ($r=\sqrt{x^2+y^2}\leq\SI{25}{\milli\meter}$), equally distributed kinetic energies between $\SI{0}{\nano\electronvolt}$ and $\SI{100}{\nano\electronvolt}$, emission angles with respect to the $z$-axis up to $\theta_{\text{max}}=\SI{25}{\degree}$ and the spin $\vec{S}$ antiparallel to $\vec{B}_z$ (HFS state). The start position in $z$-direction was set to $z = z_{\text{SF2}} - \SI{162}{\milli\meter}$ to save computing time. 
The simulation of a single neutron is aborted as soon as it is back reflected ($v_z<0$).
At $z = \SI{1100}{\milli\meter}$, the simulation is terminated and the number of successfully spin flipped neutrons $N_{\text{LFS}}$ is counted. 
The spin flip efficiency is then calculated as 
\begin{equation}\label{eq:sf_efficiency}
 \varepsilon_{\text{SF}}(\%) = \frac{N_{\text{LFS}}}{N}\cdot 100\,.
\end{equation}
The simulation was repeated with different $B_1$ amplitudes, the resulting dependency is shown in figure\,\ref{fig:sim_efficiency_vs_B1}. 
The lines are logistics functions
\begin{equation}\label{eq:logistics_function}
 \varepsilon_{\text{SF}} = \frac{\varepsilon_{\text{sat}}}{\raisebox{-0.8mm}{$1+\exp(-b(B_1-B_1^{\text{mid}}))$}}
\end{equation}
with $\varepsilon_{\text{sat}}$ the spin flip efficiency in saturation, $b$ the steepness, and $B_1^{\text{mid}}$ the amplitude with $\SI{50}{\percent}$ spin flip efficiency.
Despite lower amplitude requirements at $z_{\text{SF2},\text{a}}=\SI{980}{\milli\meter}$, in both positions, a saturation of the spin flip efficiency is achieved.
The result of this simulation supports the experimentally observed saturation behavior of the spin flip efficiency of SF2 (see section\,\ref{sec:phase_space}, figure\,\ref{fig:n_vs_power}).
The amplitude $B_1$ of the RF magnetic field is proportional to the square root of the applied RF power, $B_1 \propto \sqrt{P}$, but the proportionality constant depends on the exact tuning parameters of the LC network.
\subsection{Comparison of trap filling with two spin flippers}
\label{ssec:sim_dsf_technique}
The dependence of the effectiveness of the double spin flip technique on the beam divergence $\theta_{\text{max}}$, i.e., the maximum angle up to which neutron trajectories occur relative to the $z$-axis, was discussed in Sec.\,\ref{sec:phase_space}.
In order to compare the single and double spin flip technique with respect to $\theta_{\text{max}}$ simulations were performed with a set of neutrons ($N=4000$) with randomized positions in the $xy$-plane, a $\sqrt{E}$-dependent kinetic energy distribution between $\SI{0}{\nano\electronvolt}$ and $\SI{200}{\nano\electronvolt}$ and randomized emission angles between $\SI{0}{\degree}$ and $\theta_{\text{max}}$ with respect to the $z$-axis. 
The simulations started in front of the first field maximum of $B_z$ at $z=\SI{-1500}{\milli\meter}$. 
Both spin flipper field amplitudes were set to $B_1=\SI{0.5}{\milli\tesla}$ which is sufficient for $\varepsilon_{\text{SF}}=1$.
For the simulation of the single spin flip technique, SF1 was switched off ($B_1=\SI{0}{\milli\tesla}$). 
The time step size in the simulation was optimized in advance \cite{RossPhd}. 
\begin{figure}[b]
\centering
 \includegraphics[width=0.95\columnwidth]{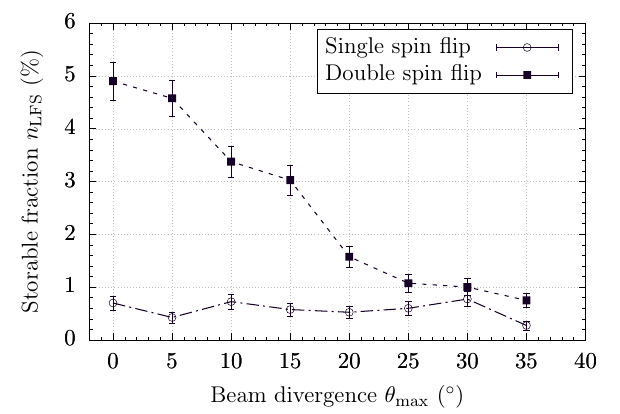}
\caption{Simulation results obtained for the single and double spin flip technique as fraction of storable neutrons $n_{\text{LFS}}$ depending on the initial beam divergence $\theta_{\text{max}}$.
The decline in $n_{\text{LFS}}$ with increasing $\theta_{\text{max}}$ is caused by a higher number of neutrons being back-reflected from the increasing magnetic potential before reaching SF2 when using the double spin flip technique.}
\label{fig:sim_doubleSF}
\end{figure}
Figure\,\ref{fig:sim_doubleSF} shows the resulting storable fraction of the initial neutron spectrum $n_{\text{LFS}} = N_{\text{LFS}}/N$ after passing the spin flippers. 
The criterion was that the neutrons must exit SF2 as low-field seekers with a total energy of $E\leq\SI{47}{\nano\electronvolt}$. 
The gravitational potential was inferred from the final $y$-position after the simulation terminated and was included in the total energy. 
It shows that up to $\theta_{\text{max}} \simeq \SI{20}{\degree}$ the double spin flip is clearly superior to the single spin flip technique. However, at higher beam divergences, both approaches become roughly comparable. 
One reason for this might be that at larger emission angles, a higher fraction of the neutrons are back reflected before they reach SF2 when the double spin-flip technique is used.
\section{Systematic effects on neutron storage time}
\label{sec:systematic effects}
The neutron storage time constant $\tau$ is related to the $\beta$-decay lifetime $\tau_\beta$ via

\begin{equation}
    \frac{1}{\tau}=\frac{1}{\tau_\beta}+\sum \frac{1}{\tau_\mathrm{X}}
\end{equation}

where $\tau_\mathrm{X}$ characterizes losses through additional mechanisms.
These systematic effects of "bottle"-type neutron lifetime experiments can be
separated into wall and non-wall related interactions.
Earlier measurements (see e.g. references 8-20 in \cite{Cude2022}) have been regularly re-analyzed and corrected for previously not known effects or shifts of effect sizes.  
Due to the fact that \tss uses fully magnetic confinement of neutrons, the first category is eliminated by design.
Material wall interactions only happen during the filling process, in preparation of the trapped neutron ensemble, before the actual storage time is started. 
Known systematic changes to the measured neutron lifetime not related to wall interactions can originate from 

\begin{enumerate}
    \item interactions of the neutrons with residual gas molecules or atoms,
    \item loss of neutron polarization due to insufficient adiabaticity of magnetic field changes,
    \item or the presence of marginally trapped neutrons in the trap after the filling process.
\end{enumerate}
1.
The superconducting coils and the permanent magnet array of \tss are situated in a cold bore cryostat.
Thus, all material surrounding the UCN storage region is cooled to approximately \SI{170}{\kelvin}, the cryostat walls even to approximately \SI{50}{\kelvin}, meaning that many gases freeze on the walls, leading to good vacuum conditions.
In steady state, the remaining pressure inside \tss is of the order of \SI{2e-7}{\milli\bar}.
This is measured on the neutron detector side of the vacuum chamber, while all pumps are connected at the source side.
Thus, the pressure inside the storage region can be assumed to be even lower. 
In contrast to historical neutron lifetime 
experiments, \tss does not use \isotope[3]{He} as a detection gas, which has a very large neutron
capture cross-section such that tiny partial pressures could lead to a significant increase of neutron losses.
To precisely determine a possible small correction to the measured lifetime, a residual gas analyzer will be connected in the future to the vacuum chamber during data taking such that scattering and absorption probabilities for the identified gases can be calculated.
2.
The magnetic field created by the superconducting solenoids and the permanent octupole magnet in \tss
is well understood. 
With the strong holding field of approximately \SI{200}{\milli\tesla}, a worst case depolarization time constant $\tau_\text{dep}=\SI{1.2e9}{\second}\gg\tau_n$ was estimated using the method described in \cite{Steyerl2017} adapted to the magnetic field conditions in \tss.
This contribution is thus negligible.
3.
Marginally trapped neutrons are neutrons on long-lived trajectories with total energies above the trapping 
potential of the magnetic field configuration. 
They can mimic a shorter neutron lifetime by leaving the trap at intermediate times, such that they are not counted after long storage times.  
The measured storage curve then shows a non-exponential behavior.
In order to prepare the UCNs inside the \tss trap such that no marginally trapped neutrons remain when the actual storage period starts, the UCN detector is inserted from its storage position $\zdetstorage$ to the cleaning position $\zdetcleaning$ further inside the trap volume. 
UCNs that can reach the boron layer of the detector at this position will be absorbed, and thus the UCN energy spectrum is cut off below the trap edge energy.
After a cleaning time $t_\text{clean}$, the detector is moved back to its storage position.
Both $z_\text{clean}$ and $t_\text{clean}$ can be optimized in order to effectively remove marginally trapped 
neutrons while efficiently keeping as many storable neutrons in the trap as possible. 
In order to make the cleaning process as effective as possible, \tss can be tilted by a small angle towards the detector side, which breaks the symmetry with respect to gravity, such that no
closed orbits exist which never intersect with the detector plane in cleaning position.
Measurements with multiple combinations of $z_\text{clean}$, $t_\text{clean}$ and tilt angle as well as detailed Monte Carlo simulations of the cleaning process are currently carried out to understand and improve the cleaning procedure.
Additional measurements and simulations are currently carried out to understand and improve the cleaning procedure.

\section{Time structure of a single measurement}
\label{sec:procedures}
\begin{figure}[b]
\centering
 \includegraphics[width=0.95\columnwidth]{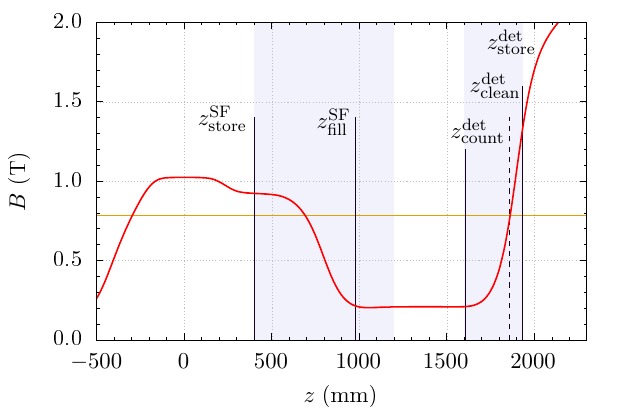}
\caption{Positions of spin flipper SF2 and detector in the longitudinal magnetic field (red line). The horizontal line (yellow) marks the trap limit at $B=\SI{0.779}{\tesla}$ imposed by $B_r^{\text{min}}(r=\SI{53.5}{\milli\meter})$. The accessible ranges of spin flipper SF2 and detector are marked by the gray boxes and range from $z=\SI{397}{\milli\meter}$ to $\SI{1200}{\milli\meter}$ for the spin flipper and $z=\SI{1599}{\milli\meter}$ to $\SI{1934}{\milli\meter}$ for the detector. The black lines indicate the positions of spin flipper and detector during individual steps of the measurement sequence with $\zsfstorage = \SI{400}{\milli\meter}$ and $\zsffill = \SI{980}{\milli\meter}$ for the spin flipper as well as $\zdetcounting = \SI{1604}{\milli\meter}$, $\zdetcleaning = \SI{1854}{\milli\meter}$ and $\zdetstorage = \SI{1932}{\milli\meter}$ for the detector.}
\label{fig:sf_and_det_positions}
\end{figure}
\begin{figure}[b]
\centering
 \includegraphics[width=0.95\columnwidth]{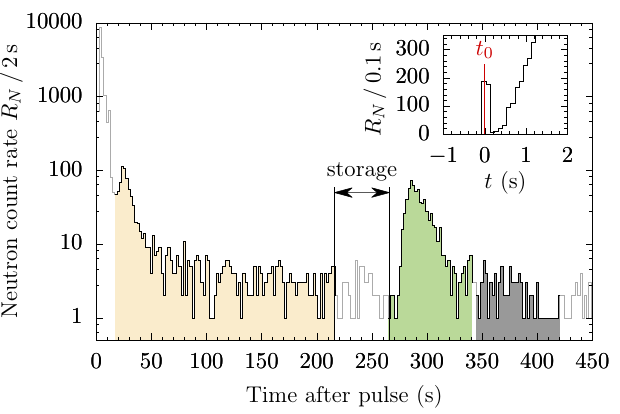}
\caption{Typical neutron detection time distribution with $\SI{200}{\second}$ cleaning duration and $\SI{50}{\second}$ storage time at $\SI{2}{\second}$ time binning. The origin of the time axis is given by the pulse of thermal neutrons, which is visible as a sharp peak in the inset ($\SI{100}{\milli\second}$ binning) and marked as $t_0$ (red line). This peak is not visible in the overview plot with larger time binning. The reactor pulse also triggers the \SI{6}{\second}-long RF pulses. Cleaning starts $\SI{10}{\second}$ after filling is finished, the yellow region marks the cleaning interval. The counting interval in $t=[265,340]\,\mathrm{s}$ is marked in green. The background interval ($t=[345,420]\,\mathrm{s}$) is shown in gray.}
\label{fig:arrival_spectrum}
\end{figure}

Each \tss measurement consists of four phases, which are controlled by a timing device providing $\SI{5}{\volt}$ and $\SI{24}{\volt}$ logic outputs.
Via these channels, all devices required to perform a measurement can be either triggered to perform predefined actions or are active as long as the corresponding output is switched on. 
Initially, the detector is in its storage position ($\zdetstorage = \SI{1932}{\milli\meter}$) and the spin flipping unit is at the filling position ($\zsffill = \SI{980}{\milli\meter}$) as in figure\,\ref{fig:sf_and_det_positions}.\\
\textbf{Filling phase}: When the reactor pulse is triggered, the SFU is switched on. It generates the rotating $B_1$ field and UCNs are filled into the trap. The optimum active duration was found to be $\fillingduration=\SI{6}{\second}$. 
Subsequently, the SFU is retracted from the trap volume to its storage position ($\zsfstorage=\SI{400}{\milli\meter}$) in the high field.\\
\textbf{Cleaning phase}: The detector is moved to the fringe of the storage volume ($\zdetcleaning=\SI{1854}{\milli\meter}$) for $\SI{200}{\second}$ in order to remove marginally trapped neutrons from the UCN spectrum (see Sec.\,\ref{sec:systematic effects}). 
Afterward, it is retracted to its storage position ($\zdetstorage$).\\
\textbf{Storage phase}:
The logic signal causing the detector stepper motor to retract the detector to its storage position marks the beginning of the nominal storage time, which is typically varied between $\SI{20}{\second}$ and $\SI{2000}{\second}$.\\
\textbf{Counting phase}:
The nominal storage time ends with another logic signal to the stepper motor of the detector, which then moves to the counting position $\zdetcounting$ in the low field and counts the remaining UCNs in the trap volume. Finally, SFU and detector are moved back to their initial positions in preparation for the next measurement.
With the detector moving at a constant velocity of $\SI{10}{\milli\meter/\second}$, the storage time for each detected UCN varies. So instead of the nominal storage time, we consider the effective storage time
\begin{equation}
 t_s^{\text{eff}}=t_s + t_s'.
\end{equation}
Here, $t_s$ is the nominal storage time and $t_s'=\SI{7.8}{\second}$ is the time in which the detector moves from $\zdetcleaning$ to $\zdetstorage$. 
For simplicity, however, we will refer to the nominal storage time for the rest of this paper. \\
The detector raw data sent by the ADC (see Sec.\,\ref{ssec:detector}) is written to hard drive during the entire measurement sequence, which allows for offline waveform analyses and e.g., retrospective optimization of the event detection algorithm.
Generally, each measurement follows a specific neutron counting pattern, as is shown in figure\,\ref{fig:arrival_spectrum}.
When the pulse is triggered, a short peak caused by thermal neutrons is visible in the detector data (see inset in figure\,\ref{fig:arrival_spectrum}), the timing of which is used as reference $t_{0} = \SI{0}{\second}$.
Many neutrons (UCNs and faster neutrons) guided along the beam tube which are too fast to be stored in the trap reach the detector within the first few seconds after the reactor pulse. 
Subsequently, ($t>\SI{10}{\second}$), after the detector has been moved to the cleaning position, marginally trapped neutrons captured during the cleaning process appear as an additional peak in the arrival spectrum.
The interval of interest for measuring the storage curve is the 'counting peak' which follows after the predefined storage time. 
The start time of this interval is calculated as $\cleaningstarttime + \cleaningduration + t_s + \SI{5}{\second}$ and it has a length of $\SI{75}{\second}$ with negligible uncertainty below \SI{0.1}{\second}. 
The length was estimated from the duration when the neutron count rate drops back to background level. 
In principle, the neutron lifetime needs to be factored in when integrating the counting peak, as both detector and neutrons move at finite velocity. 
That is, each of the neutrons experiences a different storage time and thus has a different decay probability. 
By ensuring the counting procedure has the identical time structure for every measurement, effects from differing storage times on the measured lifetime are eliminated. 
The background count rate is obtained from an interval with a length of $\SI{75}{\second}$ starting $\SI{5}{\second}$ after the end of the nominal counting interval. 
On average, the background count rate is $\sim\SI{1.1}{\hertz}$.
The number of counted neutrons in the background interval is subsequently subtracted from the number of counted neutrons in the counting interval.
\section{First UCN storage curve}
\label{sec:storage_curve}
\begin{figure*}[t]
\centering
 \includegraphics[width=0.95\textwidth]{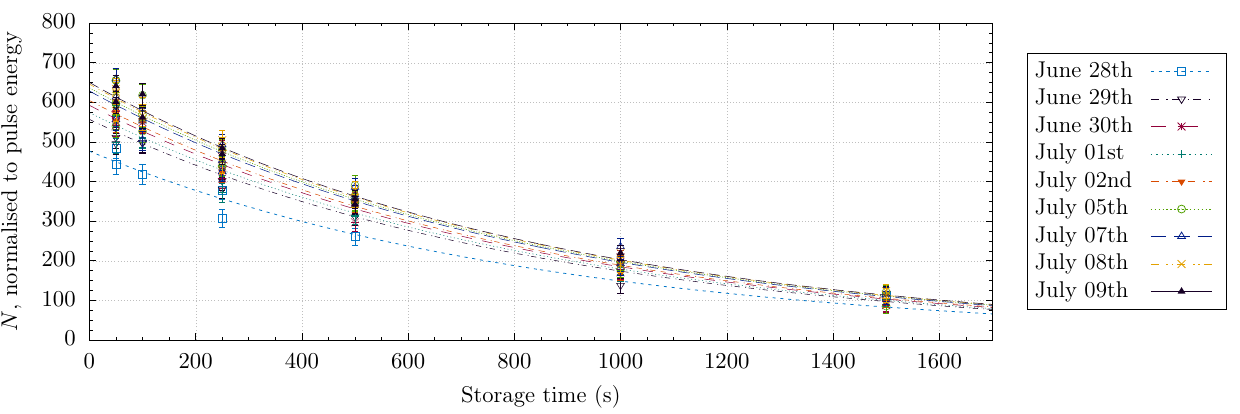}
\caption{Storage curve measured in June/July 2021 at the research reactor TRIGA Mainz with cleaning for $\SI{200}{\second}$ at $\zdetcleaning=\SI{1854}{\milli\meter}$. The lines are a multidimensional fit (\ref{eq:multibranch}) with each day assigned a separate amplitude but a common decay constant $\tau = \SI{859\pm16}{\second}$. The remaining fit parameters are listed in table\,\ref{tab:sc_cleaning_200s}.}
\label{fig:storage_curve_200s_cleaning}
\end{figure*}
After first parameter optimizations of the filling and cleaning procedures, a storage curve was measured in June/July 2021 (see figure\,\ref{fig:storage_curve_200s_cleaning}). 
Cleaning was done at $\zdetcleaning=\SI{1854}{\milli\meter}$ for $\SI{200}{\second}$. 
During the beamtime strong daily fluctuations of the UCN source were observed. 
To account for this observation the data were split by days and a multidimensional fit of the form
\begin{equation}\label{eq:multibranch}
 N(t) = N_0^i\exp (-t/\tau)
\end{equation}
was applied, allowing for varying initial numbers of neutrons, $N_0^i$, but using the same decay constant $\tau$. 
The fit parameters are listed in Table \,\ref{tab:sc_cleaning_200s}. 
Our first estimation for the neutron storage time constant is 
\begin{equation}
 \tau = \SI{859 \pm 16}{\second} .
\end{equation}
We note that this value is smaller than the current particle data group (PDG 2022) average of $\SI{878.4\pm0.5}{\second}$ \cite{PDG2022}. 
However, we point out that final systematic studies and corresponding optimizations of for example the parameters of the cleaning process as described in Sec.\,\ref{sec:systematic effects}, have not yet concluded, such that this value should not be interpreted as a measurement of the free neutron lifetime.
%
%
%
To increase the counting statistics, we will further investigate optimization of the filling process using the double spin flip technique. 
Additionally, systematic investigations are ongoing, which include Monte Carlo simulations of the cleaning process.
\begin{table}
\centering
\caption{Multidimensional fit results of a single exponential function (cf. \ref{eq:multibranch}) to the storage curve depicted in figure\,\ref{fig:storage_curve_200s_cleaning}.}
\label{tab:sc_cleaning_200s}       
\begin{tabular}{lll}
\hline\noalign{\smallskip}
Measurement day & Parameter & Value  \\
\noalign{\smallskip}\hline\noalign{\smallskip}
June 28th & $N_0^1$ & $477(13)$\\
  June 29th & $N_0^2$ & $557(11)$\\
  June 30th & $N_0^3$ & $592(11)$\\
  July 01st & $N_0^4$ & $574(10)$\\
  July 02nd & $N_0^5$ & $605(15)$\\
  July 05th & $N_0^6$ & $636(10)$\\
  July 07th & $N_0^7$ & $628(11)$\\
  July 08th & $N_0^8$ & $646(10)$\\
  July 09th & $N_0^9$ & $650(13)$\\\hline
 & $\tau$ & $\SI{859(16)}{\second}$\\\hline
 & $\chi^2/\text{ndf}$ & $1.14$\\
\noalign{\smallskip}\hline
\end{tabular}
\end{table}
\section{Summary}
\label{sec:summary}
The \tss experiment has for the first time demonstrated the loading of a three-dimensional magnetic field gradient trap with externally-produced UCNs using a pulsed spin-flip loading scheme. 
The implementation of a magnetic field gradient compensation octupole in combination with adiabatic fast passage spin flippers allows for trap loading with subsequent retraction of all material surfaces from the storage volume. 
A sequential implementation of spin flippers increases and shifts the kinetic energy acceptance interval of the \tss experiment into a region of higher production rate from the UCN source. 
UCN losses related to absorption on residual gas or due to spin-depolarization have been shown to be negligible for a future measurement of the free neutron lifetime.
Our custom-built, SiPM-based in-situ UCN detector serves a dual purpose. 
Partially inserted into the trapping volume, it absorbs marginally trapped neutrons with kinetic energies above the nominal trap depth. 
The parameter space for this spectral cleaning procedure was investigated within the available statistics. 
After the preset neutron storage time, the detector is fully inserted into the trap volume to count the remaining UCNs without additional transport losses. 
Operation of \tss at the research reactor TRIGA Mainz allowed to extract a neutron storage time constant of $\tau=\SI{859(16)}{s}$, compatible with the most precise, but not to be interpreted as a  measurement of the free neutron lifetime.
The complete relocation of the \tss experiment in 2023 to the beam port West 1 of the UCN source of the Paul-Scherrer-Institute will allow the investigation of all relevant systematic effects required to perform an independent measurement of the neutron lifetime to contribute to the investigation of the neutron lifetime puzzle.  
These measurements will be used to design an experiment that can contribute to a competitive determination of \Vud in the future. 

\ack
We thank the administration of the research reactor TRIGA Mainz as well as the reactor operators for providing us beamtime. We are grateful to the mechanical and electrical workshops of the Department of Chemistry and of the Institute of Physics at JGU Mainz.
We thank S. Vanneste for his careful reading of the text and helpful comments.
The development of \tss has been supported by the Cluster of Excellence “Precision Physics, Fundamental Interactions, and Structure of Matter” (PRISMA+ EXC 2118/1) funded by the German Research Foundation (DFG) within the German Excellence Strategy (Project ID 39083149).
%
%
\section*{References}
\bibliographystyle{spphys.bst}       
\bibliography{tauSpect}   
\end{document}